\documentclass[3p, times]{elsarticle}
\usepackage{graphicx}

\usepackage{amssymb}

\begin{document}

\begin{frontmatter}




\title{Pinning down QCD-matter shear viscosity in A+A collisions via EbyE fluctuations using pQCD + saturation + hydrodynamics}


\author[a]{H.~Niemi}
\author[b,c]{K.~J.~Eskola}
\author[d]{R.~Paatelainen}
\author[c,e]{K.~Tuominen}

\address[a]{Institut f\"ur Theoretische Physik, Johann Wolfgang Goethe-Universit\"at, Max-von-Laue-Str. 1, D-60438 Frankfurt am Main, Germany}
\address[b]{University of Jyv\"askyl\"a, Department of Physics, P.O.~Box 35, FI-40014 University of Jyv\"askyl\"a, Finland}
\address[c]{Helsinki Institute of Physics, P.O.Box 64, FI-00014 University of Helsinki, Finland}
\address[d]{Departamento de Fisica de Particulas, Universidade de Santiago de Compostela, E-15782 Santiago de Compostela, Galicia, Spain} 
\address[e]{Department of Physics, University of Helsinki, P.O.~Box 64, FI-00014 University of Helsinki, Finland}



\begin{abstract}
We compute the initial energy densities produced in ultrarelativistic heavy-ion collisions from NLO perturbative QCD using a saturation conjecture to control soft particle production, and describe the subsequent space-time evolution of the system with hydrodynamics, event by event. The resulting centrality dependence of the low-$p_T$ observables from this pQCD + saturation + hydro ("EKRT") framework are then compared simultaneously to the LHC and RHIC measurements. With such an analysis we can test the initial state calculation, and constrain the temperature dependence of the shear viscosity-to-entropy ratio $\eta/s$ of QCD matter. Using these constraints from the current RHIC and LHC measurements we then predict the charged hadron multiplicities and flow coefficients for the 5.023 TeV Pb+Pb collisions.
\end{abstract}




\end{frontmatter}


%
\section{Introduction}
\label{sec:intro}

The basic assumption in the EKRT framework is that the initial transverse energy production in ultrarelativistic heavy-ion collisions can be calculated by using perturbative QCD and collinear factorization, and that the production is locally controlled by a semi-hard saturation scale $p_{\rm sat}$ \cite{Eskola:1999fc, Paatelainen:2012at}. 

Using hydrodynamics with the EKRT initial conditions, we show that we can describe a multitude of low-$p_T$ observables simultaneously at RHIC and LHC energies, and constrain the QCD $\eta/s(T)$ \cite{Niemi:2015qia}. Moreover, once the model parameters are fixed at one collision energy, the results for the other collision energies are predictions, and we can predict the low-$p_T$ observables in 5.023 TeV Pb+Pb collisions at the LHC \cite{Niemi:2015voa}.

The minijet transverse energy $E_T$ produced into the rapidity interval $\Delta y$ per unit transverse area in $A+A$ collisions can be computed by using collinear factorization as \cite{Paatelainen:2012at}
\begin{equation}
\frac{dE_T}{d^2{\bf r}}(p_0, \sqrt{s_{NN}}, A, \Delta y, \mathbf{r}, \mathbf{b}; \beta) = T_A(\mathbf{r}+ \mathbf{b}/2)T_A(\mathbf{r}- \mathbf{b}/2)\sigma\langle E_T \rangle_{p_0,\Delta y,\beta},
\label{eq: dET}
\end{equation}
where $p_0\gg \Lambda_{\textrm{QCD}}$ is the transverse momentum cut-off scale, $\mathbf{r}$ is the transverse coordinate and $\mathbf{b}$ is the impact parameter and $T_A$ is the nuclear thickness function. The basic input to the calculation of $\sigma\langle E_T \rangle_{p_0,\Delta y,\beta}$ are the NLO pQCD partonic cross-sections \cite{Ellis:1985er,Paatelainen:2014fsa} and the nuclear parton distributions \cite{Helenius:2012wd}. The parameter $\beta \in [0,1]$ (here $\beta = 0.8$) controls the minimum $E_T$ in $\Delta y$ defined in the measurement functions that render the NLO calculation of $E_T$ infra-red and collinear safe \cite{Paatelainen:2012at}. 

The cut-off scale $p_0$ is obtained from the NLO generalization \cite{Paatelainen:2012at} of the EKRT saturation condition \cite{Eskola:1999fc}
\begin{equation}
\frac{\mathrm{d}E_T}{\mathrm{d}^2\mathbf{r}}(p_0,\sqrt{s_{NN}},A,\Delta y,\mathbf{r},\mathbf{b};\beta) = \frac{K_{\rm sat}}{\pi}p_0^3\Delta y,
\label{eq:satcri}
\end{equation}
where $K_{\rm sat}$ is a free parameter. The $\mathbf{r}$ dependence of the solution $p_0 = p_{\rm sat}(\sqrt{s_{NN}},A,\Delta y,\mathbf{r},\mathbf{b};\beta,K_{\rm sat})$ enters essentially only through $T_A(\mathbf{r})$ \cite{Paatelainen:2013eea,Eskola:2001rx}, i.e. $p_{\rm sat} \sim [T_A T_A]^n$. 

The event-by-event fluctuations enter through the fluctuations in $T_A$. These are calculated by first sampling the nucleon positions from the Woods-Saxon nucleon density profiles, and then setting a gaussian transverse density around each nucleon, with width $\sigma = 0.43$ fm, obtained from the measurements of $J/\psi$ electroproduction \cite{Chekanov:2004mw}. The nuclear $T_A$ is then a sum of the nucleon thickness functions.

The local energy density at the formation time $\tau_s(\mathbf{r}) = 1/p_{\rm sat}(\mathbf{r})$ is then
\begin{equation}
e(\mathbf{r},\tau_{\mathrm{s}}(\mathbf{r})) = \frac{\mathrm{d}E_T}{\mathrm{d}^2\mathbf{r}}\frac{1}{\tau_{\mathrm{s}} (\mathbf{r}) \Delta y } = \frac{K_{\rm sat}}{\pi}[p_{\rm sat}(\mathbf{r})]^4.
\end{equation}
The energy density profile need still to be evolved to a common time $\tau_0 =1/p_{\rm sat}^{\rm min}=0.2$~fm, at which we start the hydro. We use simple 0+1 D Bjorken hydrodynamics for this ``pre-thermal'' evolution. At the edges of the system, where  $p_{\rm sat}^{\rm min} = 1$ GeV, the energy density profile is connected smoothly to a binary profile.

\begin{figure}
\begin{center}
\includegraphics[width=6.3cm]{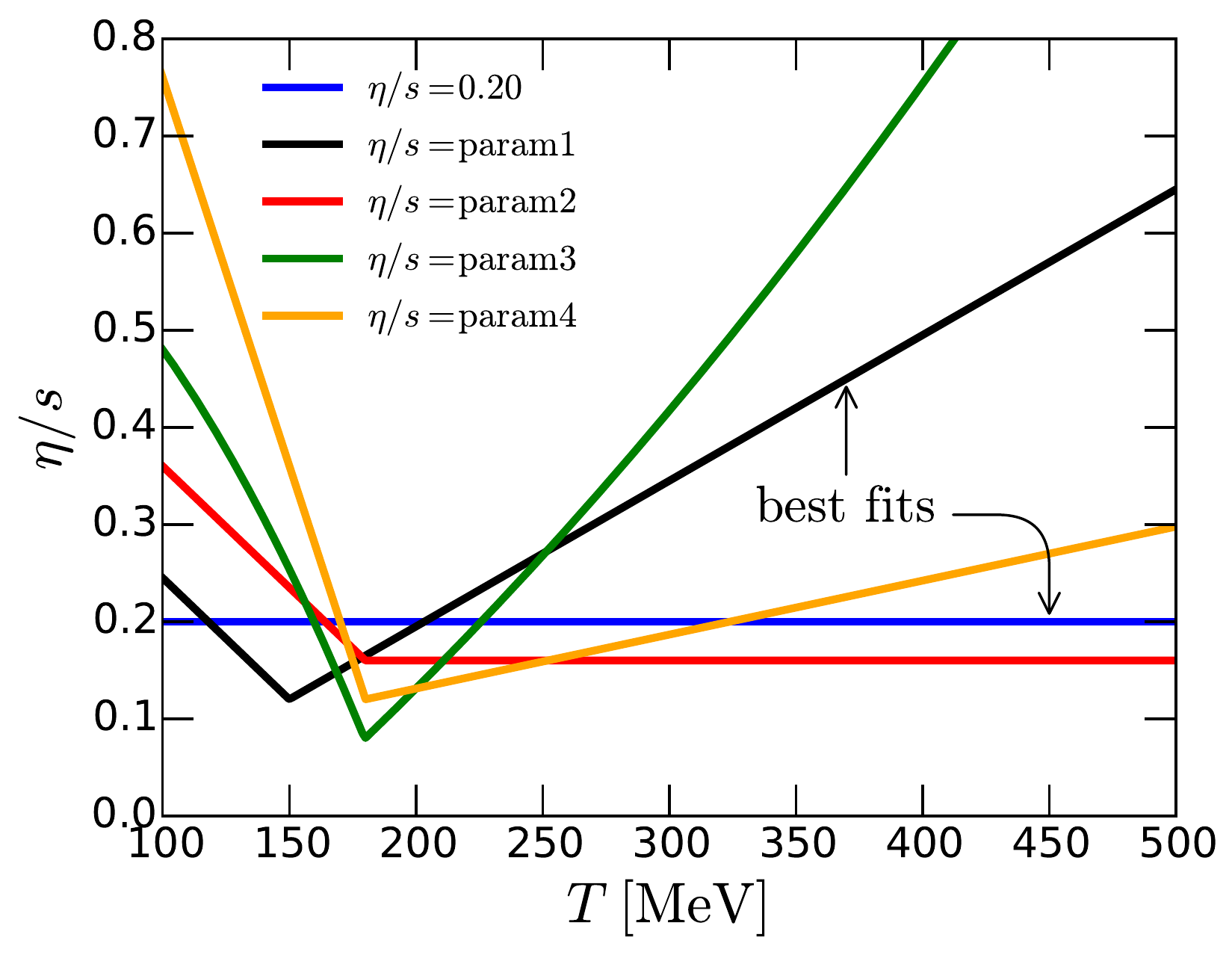}
\end{center}
\vspace{-0.5cm}
\caption{\small The tested temperature dependences of $\eta/s$. From \cite{Niemi:2015voa}.}
\vspace{-0.3cm}
\label{fig:etapers}
\end{figure}

The spacetime evolution of the system is then solved by 2+1 D dissipative fluid dynamics, with the coefficients of the second-order terms in the equations of motion from Refs.~\cite{Denicol:2012cn, Molnar:2013lta}. The equation of state is the $s95p$-PCE-v1 parametrization \cite{Huovinen:2009yb} with chemical decoupling at $T_{\rm chem} = 175$ MeV. The kinetic freeze-out is at $T_{\rm dec}=100$ MeV. The transverse flow and the components of the shear-stress tensor are initially set to zero. 
We neglect heat conductivity and bulk viscosity. The remaining hydrodynamic input is then $\eta/s(T)$. The parametrizations used in this work \cite{Niemi:2015qia, Niemi:2015voa} are shown in Fig.~\ref{fig:etapers}.

\section{Results}
\label{sec:results}

\begin{figure}
\begin{center}
\includegraphics[width=7.3cm]{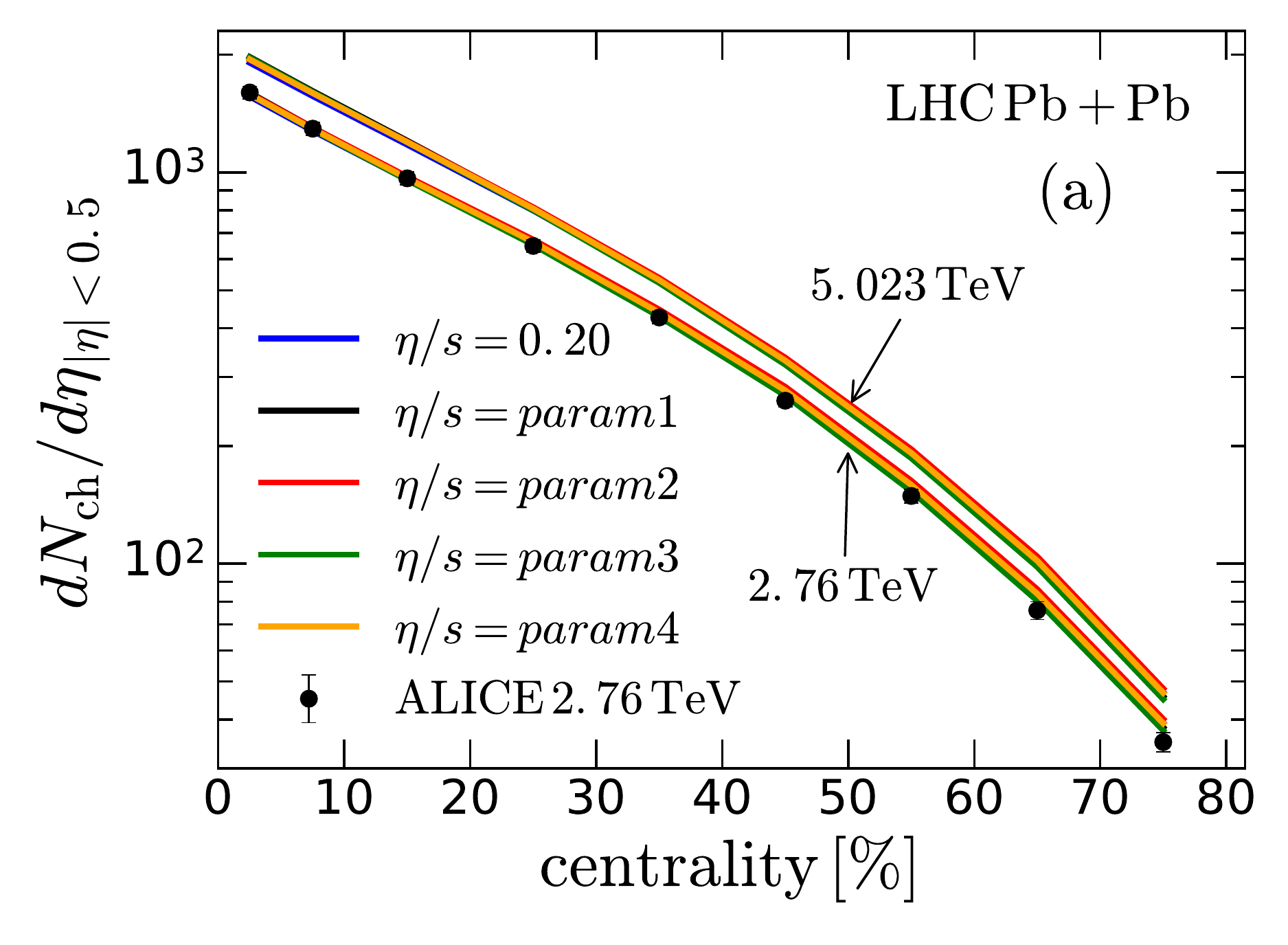}
\includegraphics[width=7.3cm]{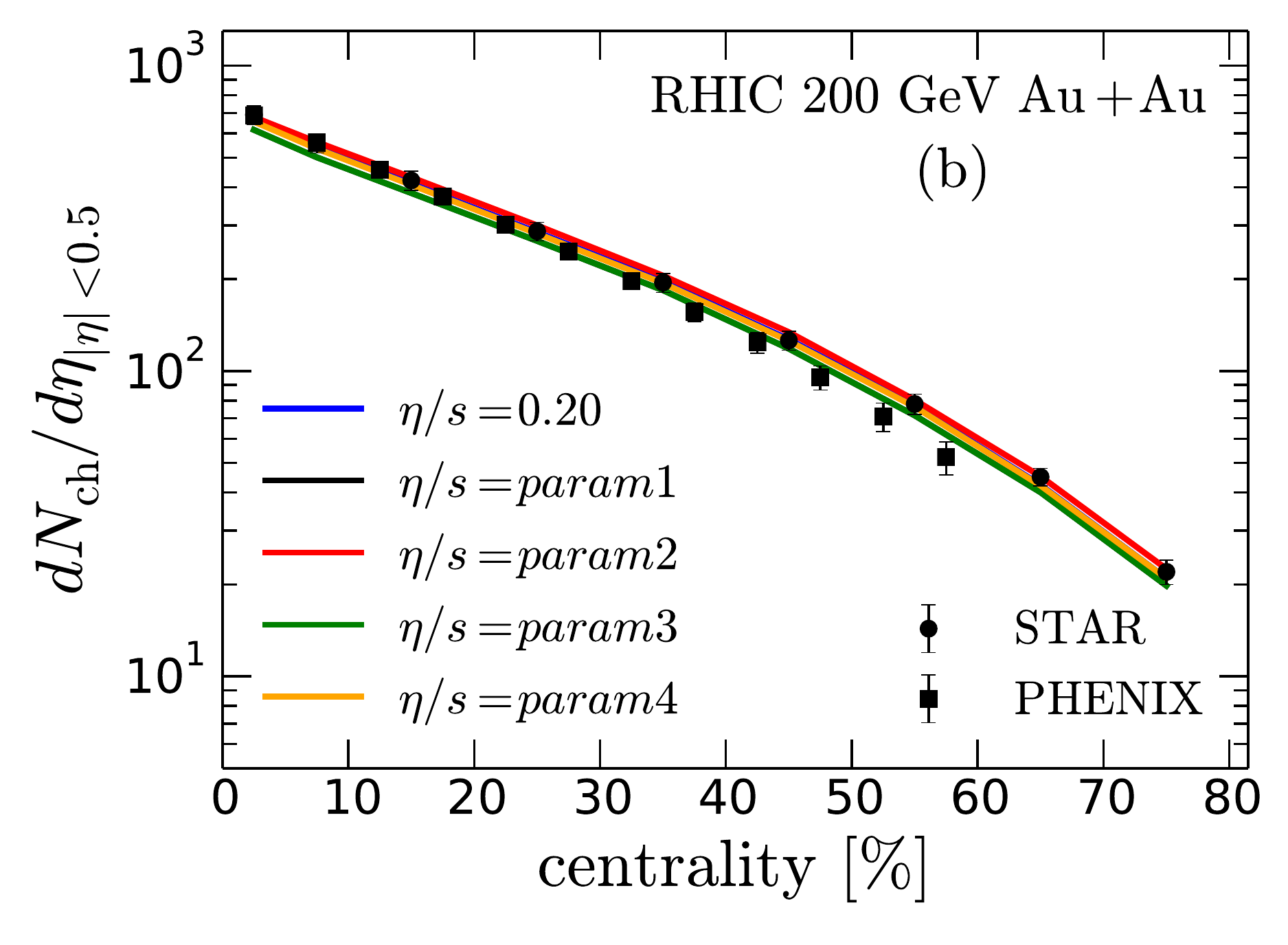}
\end{center}
\vspace{-0.5cm}
\caption{\small Charged hadron multiplicity in 2.76 TeV and 5.023 TeV Pb+Pb collisions (Left) and in 200 GeV Au+Au collisions (Right). Experimental data are from ALICE \cite{Aamodt:2010cz}, STAR \cite{Abelev:2008ab} and PHENIX \cite{Adler:2004zn}. From \cite{Niemi:2015voa} and \cite{Niemi:2015qia}. }
\vspace{-0.3cm}
\label{fig:multiplicity}
\end{figure}

Figure \ref{fig:multiplicity}a shows the centrality dependence of the charged hadron multiplicity in 2.76 TeV Pb+Pb collisions at the LHC. The measurement of the multiplicity in $0-5$ \% most central collisions is used to fix the proportionality constant $K_{\rm sat}$. The studied $\eta/s(T)$ parametrizations lead to different entropy production during the evolution, and therefore $K_{\rm sat}$ must be fixed separately for each $\eta/s(T)$. However, once $K_{\rm sat}$ is fixed, the centrality dependence of the multiplicity, as well as its dependence on the collision energy, is a prediction of the model. The calculated multiplicities in 200 GeV Au+Au collisions at RHIC are shown in Fig.~\ref{fig:multiplicity}b, and the prediction for the 5.023 TeV Pb+Pb collisions is shown by the upper set of curves in Fig.~\ref{fig:multiplicity}a. 

\begin{figure}[h]
\begin{center}
\includegraphics[width=7.4cm]{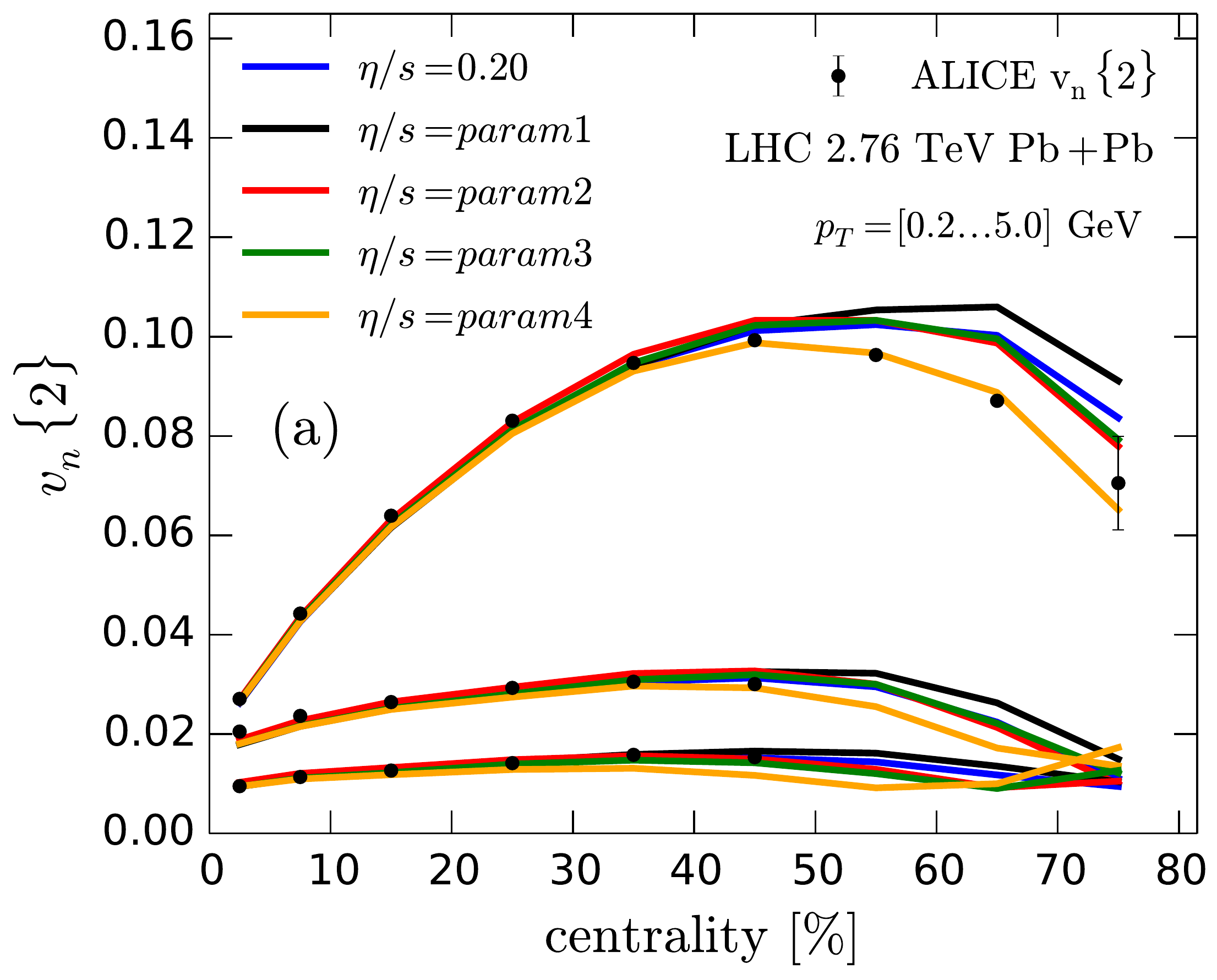}
\includegraphics[width=7.4cm]{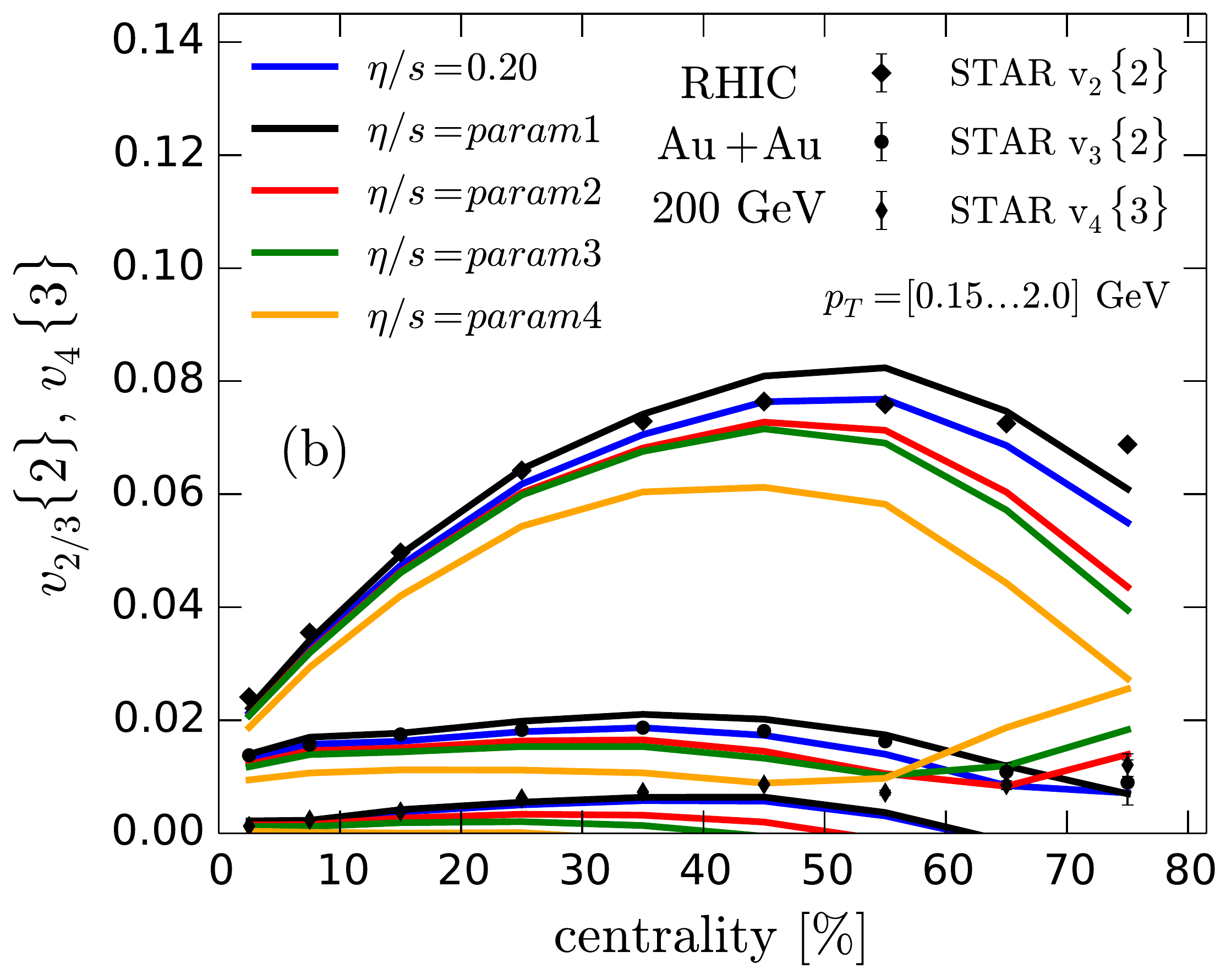}
\end{center}
\vspace{-0.7cm}
\caption{\small Flow coefficients in 2.76 TeV collisions (Left) and in 200 GeV Au+Au collisions (Right). Experimental data are from ALICE \cite{ALICE:2011ab} and STAR \cite{Adams:2004bi, Adamczyk:2013waa, Adams:2003zg}. From \cite{Niemi:2015qia}.}
\label{fig:vn}
\end{figure}

Each $\eta/s(T)$ parametrization is constructed in a such way that it gives a good description of the elliptic flow measured in mid-peripheral LHC 2.76 Tev Pb+Pb collisions. The centrality dependence of the $v_n\{2\}$ is shown in Fig.~\ref{fig:vn}a, compared to the ALICE measurements \cite{ALICE:2011ab}. As can be seen from the figure, the centrality dependence of the flow coefficients at the LHC can be described by many different temperature dependencies of the viscosity. Tighter constraints for $\eta/s(T)$ can be obtained by a simultaneous analysis of 200 GeV and 2.76 TeV collisions. This is shown in Fig.~\ref{fig:vn}b, where the EKRT results, with the same $\eta/s(T)$ parametrizations that describe the LHC data, are compared to the STAR measurements~\cite{Adams:2004bi, Adamczyk:2013waa, Adams:2003zg}.   

\begin{figure}
\begin{center}
\includegraphics[width=7.57cm]{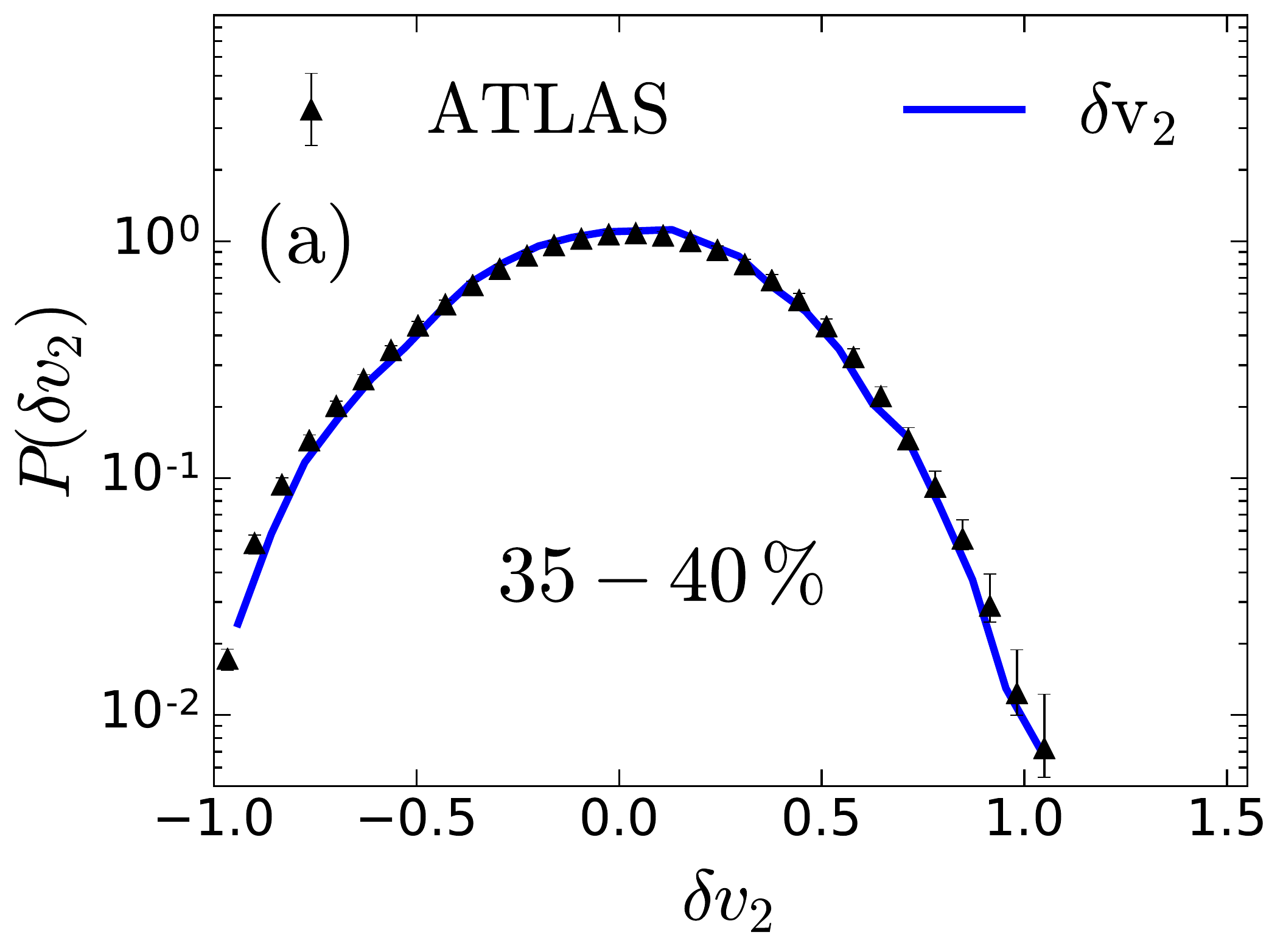}
\includegraphics[width=7.4cm]{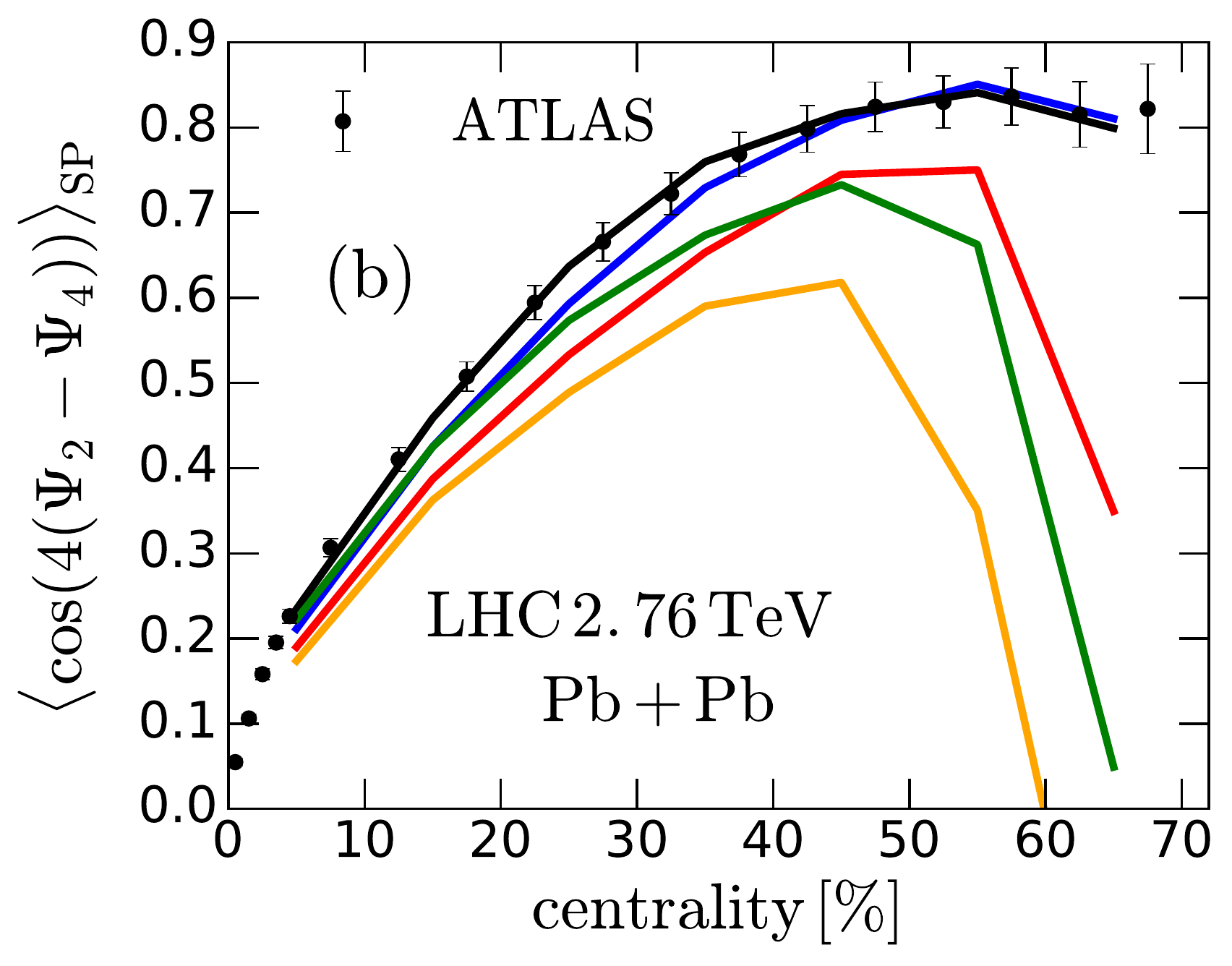}
\end{center}
\vspace{-0.7cm}
\caption{\small Elliptic flow fluctuations (Left) and event-plane angle correlations (Right) in 2.76 TeV Pb+Pb collisions. The experimental data are from ATLAS \cite{Aad:2013xma, Aad:2014fla}. From \cite{Niemi:2015qia}.}
\label{fig:fluctuations}
\end{figure}

The relative fluctuation spectra of elliptic flow, $\delta v_2 = (v_2 - \langle v_2\rangle_{\rm ev})/\langle v_2\rangle_{\rm ev}$, provide more direct constraints to the initial conditions, as they are independent of the viscosity \cite{Niemi:2015qia}. The calculated probability distribution $P(\delta v_n)$ is shown in Fig.~\ref{fig:fluctuations} together with the ATLAS data~\cite{Aad:2013xma}. 

\begin{figure}
\begin{center}
\includegraphics[width=14.1cm]{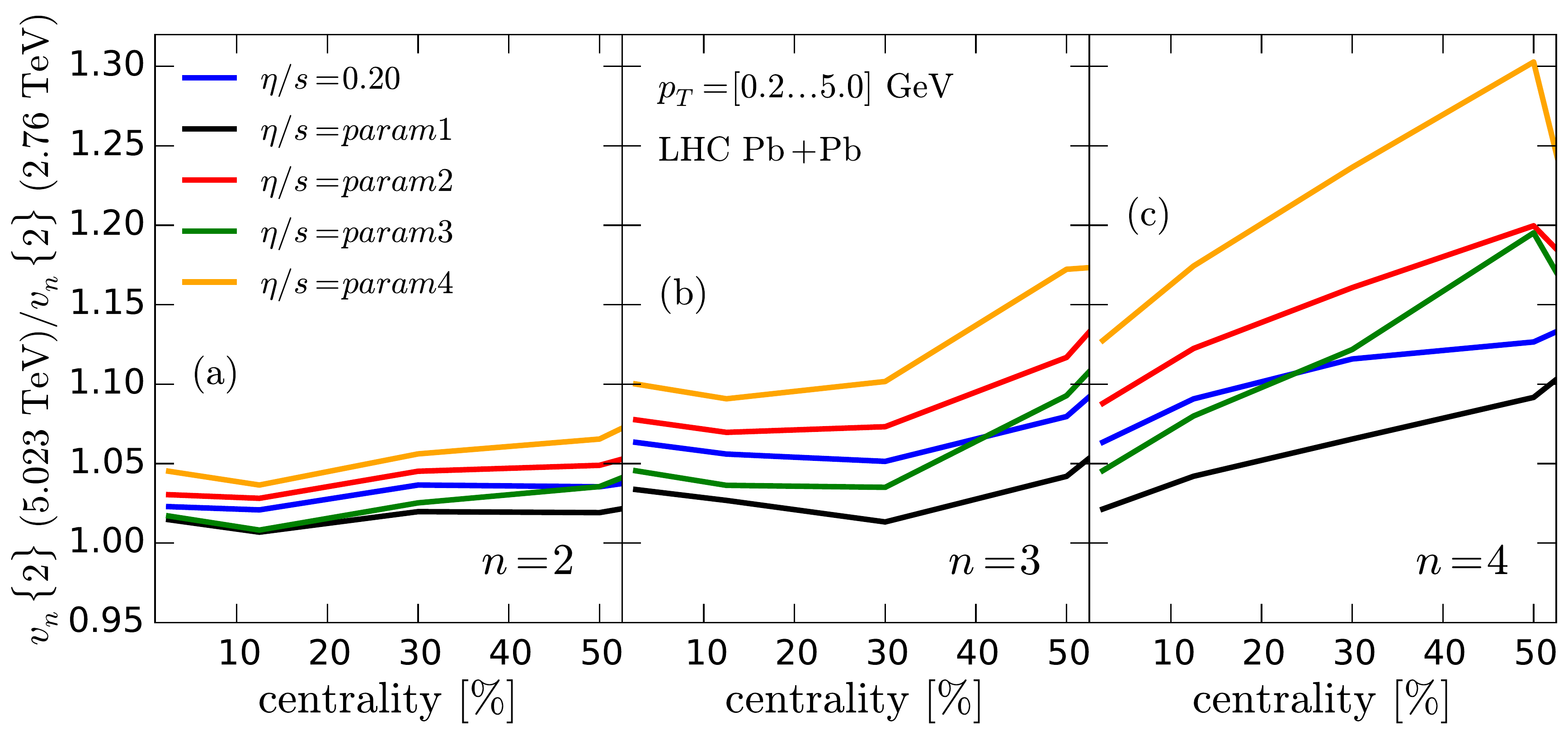}
\end{center}
\vspace{-0.5cm}
\caption{\small  Ratio of the flow coefficients $v_n\{2\}$ in 5.023 TeV and 2.76 TeV Pb+Pb collisions at the LHC. From \cite{Niemi:2015voa}.}
\label{fig:vn_prediction}
\end{figure}

Besides the flow coefficients themselves, the correlations between the event-plane angles provide vital additional constraints to $\eta/s(T)$. An example of such correlations is shown in Fig.~\ref{fig:fluctuations}b, where the EKRT model calculation with different $\eta/s(T)$ parametrizations is compared to the ATLAS data~\cite{Aad:2014fla}. It is remarkable that the same $\eta/s(T)$ parametrizations, $\eta/s=0.2$ and $param1$, that give the best description of the flow coefficients at RHIC, also give the best description of the event-plane correlations at the LHC.
Finally, in Fig.~\ref{fig:vn_prediction} we show the EKRT prediction for the ratio of the 5.023 TeV and 2.76 TeV flow coefficients.






\begin{thebibliography}{00}



\bibitem{Eskola:1999fc} 
  K.~J.~Eskola, K.~Kajantie, P.~V.~Ruuskanen and K.~Tuominen,
  Nucl.\ Phys.\ B {\bf 570}, 379 (2000)
  [hep-ph/9909456].

\bibitem{Paatelainen:2012at} 
  R.~Paatelainen, K.~J.~Eskola, H.~Holopainen and K.~Tuominen,
  Phys.\ Rev.\ C {\bf 87}, no. 4, 044904 (2013)
  [arXiv:1211.0461 [hep-ph]].

\bibitem{Niemi:2015qia} 
  H.~Niemi, K.~J.~Eskola and R.~Paatelainen,
  arXiv:1505.02677 [hep-ph], submitted to Phys. Rev. C.

\bibitem{Niemi:2015voa} 
  H.~Niemi, K.~J.~Eskola, R.~Paatelainen and K.~Tuominen,
  arXiv:1511.04296 [hep-ph], submitted to Phys. Rev. C.
  
\bibitem{Ellis:1985er} 
  R.~K.~Ellis and J.~C.~Sexton,
  Nucl.\ Phys.\ B {\bf 269}, 445 (1986).
  
\bibitem{Paatelainen:2014fsa} 
  R.~Paatelainen,
  PhD thesis,
  arXiv:1409.3508 [hep-ph].
  
\bibitem{Helenius:2012wd} 
  I.~Helenius, K.~J.~Eskola, H.~Honkanen and C.~A.~Salgado,
  JHEP {\bf 1207}, 073 (2012)
  [arXiv:1205.5359 [hep-ph]].

\bibitem{Paatelainen:2013eea} 
  R.~Paatelainen, K.~J.~Eskola, H.~Niemi and K.~Tuominen,
  Phys.\ Lett.\ B {\bf 731}, 126 (2014)
  [arXiv:1310.3105 [hep-ph]].

\bibitem{Eskola:2001rx} 
  K.~J.~Eskola, K.~Kajantie and K.~Tuominen,
  Nucl.\ Phys.\ A {\bf 700}, 509 (2002)
  [hep-ph/0106330].
  
\bibitem{Chekanov:2004mw} 
  S.~Chekanov {\it et al.} [ZEUS Collaboration],
  Nucl.\ Phys.\ B {\bf 695}, 3 (2004)
  [hep-ex/0404008].

\bibitem{Denicol:2012cn} 
  G.~S.~Denicol, H.~Niemi, E.~Molnar and D.~H.~Rischke,
  Phys.\ Rev.\ D {\bf 85}, 114047 (2012)
  [Phys.\ Rev.\ D {\bf 91}, no. 3, 039902 (2015)]
  [arXiv:1202.4551 [nucl-th]].
  
\bibitem{Molnar:2013lta} 
  E.~Moln\'ar, H.~Niemi, G.~S.~Denicol and D.~H.~Rischke,
  Phys.\ Rev.\ D {\bf 89}, no. 7, 074010 (2014)
  [arXiv:1308.0785 [nucl-th]].

\bibitem{Huovinen:2009yb} 
  P.~Huovinen and P.~Petreczky,
  Nucl.\ Phys.\ A {\bf 837}, 26 (2010)
  [arXiv:0912.2541 [hep-ph]].

\bibitem{Aamodt:2010cz} 
  K.~Aamodt {\it et al.} [ALICE Collaboration],
  Phys.\ Rev.\ Lett.\  {\bf 106}, 032301 (2011)
  [arXiv:1012.1657 [nucl-ex]].

\bibitem{Abelev:2008ab} 
  B.~I.~Abelev {\it et al.} [STAR Collaboration],
  Phys.\ Rev.\ C {\bf 79}, 034909 (2009)
  [arXiv:0808.2041 [nucl-ex]].

\bibitem{Adler:2004zn} 
  S.~S.~Adler {\it et al.} [PHENIX Collaboration],
  Phys.\ Rev.\ C {\bf 71}, 034908 (2005)
  [Phys.\ Rev.\ C {\bf 71}, 049901 (2005)]
  [nucl-ex/0409015].
  
\bibitem{ALICE:2011ab} 
  K.~Aamodt {\it et al.} [ALICE Collaboration],
  Phys.\ Rev.\ Lett.\  {\bf 107}, 032301 (2011)
  [arXiv:1105.3865 [nucl-ex]].

\bibitem{Adams:2004bi} 
  J.~Adams {\it et al.} [STAR Collaboration],
  Phys.\ Rev.\ C {\bf 72}, 014904 (2005)
  [nucl-ex/0409033].
  
\bibitem{Adamczyk:2013waa} 
  L.~Adamczyk {\it et al.} [STAR Collaboration],
  Phys.\ Rev.\ C {\bf 88}, no. 1, 014904 (2013)
  [arXiv:1301.2187 [nucl-ex]].
  
\bibitem{Adams:2003zg} 
  J.~Adams {\it et al.} [STAR Collaboration],
  Phys.\ Rev.\ Lett.\  {\bf 92}, 062301 (2004)
  [nucl-ex/0310029].
  
  
\bibitem{Aad:2013xma} 
  G.~Aad {\it et al.} [ATLAS Collaboration],
  JHEP {\bf 1311}, 183 (2013)
  [arXiv:1305.2942 [hep-ex]].

\bibitem{Aad:2014fla} 
  G.~Aad {\it et al.} [ATLAS Collaboration],
  Phys.\ Rev.\ C {\bf 90}, no. 2, 024905 (2014)
  [arXiv:1403.0489 [hep-ex]].


\end{thebibliography}




\end{document}